\documentclass[10pt]{wiley2sp}
\usepackage{url}
\usepackage{helvet}
\usepackage{psfig}
\usepackage{times}
\def\sectcounterend{.}
\def\figurename{Figure}

\def\pages{1--5}
\def\thevol{1}
\def\thenumber{1}
\def\theyear{2004}
\def\thedoi{20040001}

\begin{document}
\titlefigure[clip,width=.8\columnwidth]{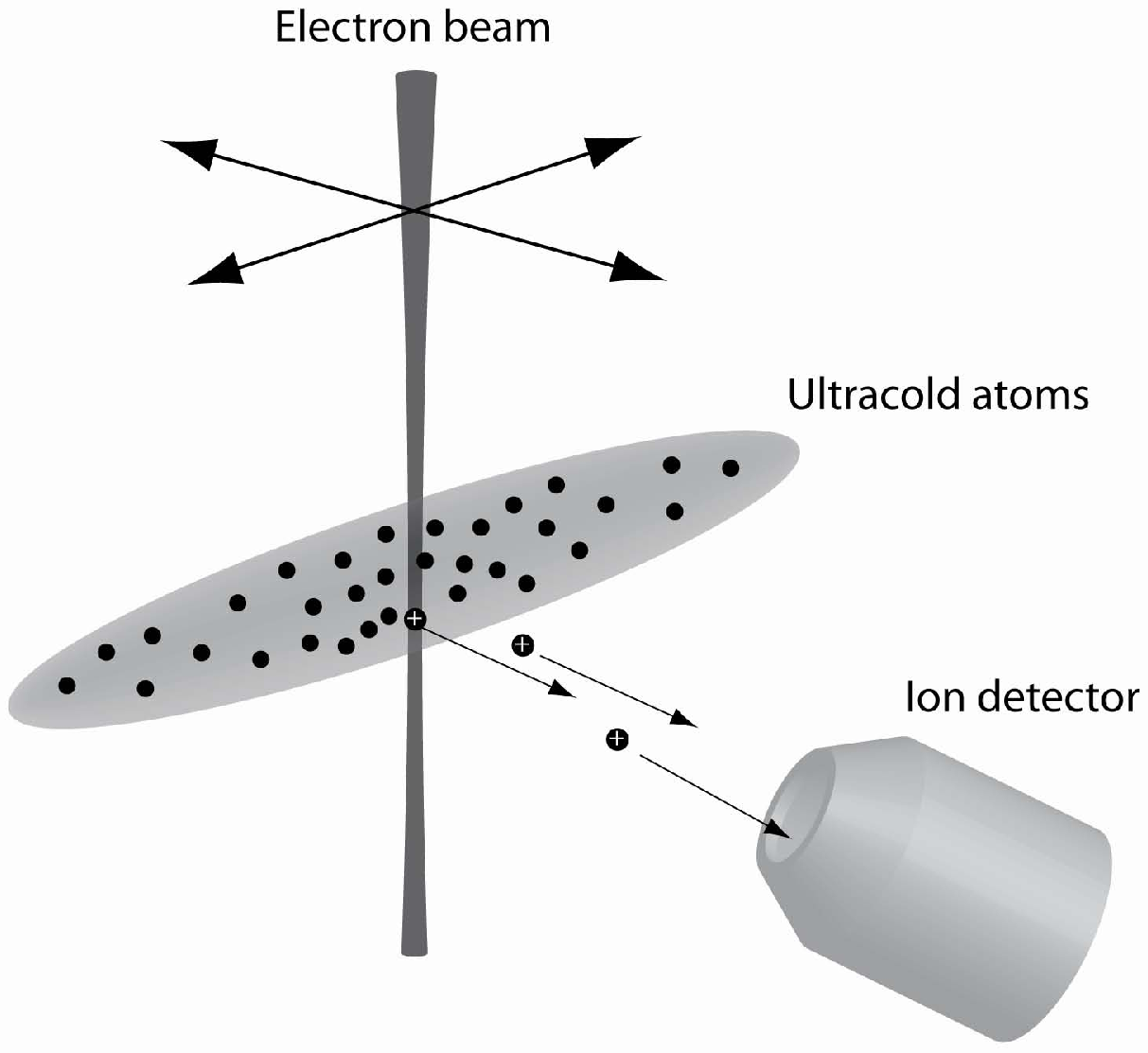}

\abstract{
We propose a new technique for the detection of single atoms in ultracold quantum gases. The technique is based on scanning electron microscopy and employs the electron impact ionization of trapped atoms with a focussed electron probe. Subsequent detection of the resulting ions allows for the reconstruction of the atoms position. This technique is expected to achieve a much better spatial resolution compared to any optical detection method. In combination with the sensitivity to single atoms, it makes new \textit{in situ} measurements of atomic correlations possible. The detection principle is also well suited for the addressing of individual sites in optical lattices.}

\titlefigurecaption{Working principle of the electron microscope: A focussed electron beam is directed onto the ultracold atoms. The atoms are ionized by electron impact and the resulting ions are detected. 
}

\title{A Scanning Electron Microscope for Ultracold Atoms}
\addtocontents{toc}{Scope Names}{}

\author{T. Gericke, C. Utfeld, N. Hommerstad, and H. Ott\inst{*}}
\institute{Institut f{\"u}r Physik, Johannes Gutenberg Universit\"at, 55099 Mainz, Germany}
\mail{e-mail: ott@uni-mainz.de}
\received{}

\keywords{ultracold quantum gases; single atom detection; scanning electron microscopy; electron-atom scattering}

\titlerunning{A Scanning Electron Microscope for Ultracold Atoms}
\authorrunning{T. Gericke, C. Utfeld, N. Hommerstad, and H. Ott}
\pacs{03.75.Hh, 07.77.-n, 07.79.-v, 34.80.Dp}
\published{}

\maketitle
\markboth{thedoi}{thedoi}
\sloppy

\section{Introduction}
The research field of ultracold quantum gases is characterized by a rapid experimental progress. New preparation techniques have always opened new testing grounds for theoretical predictions. Some of the most important technical developments during the last years are the advent of optical dipole traps and optical lattices \cite{Stamper-Kurn1998,Barrett2001,Cennini2003,Kozuma1999,Cataliotti2001,Roati2003,Greiner2002}, the realization of atom chips \cite{Ott2001,Reichel2001} and the use of magnetic fields to control the interatomic interaction via Feshbach resonances \cite{Vuletic1999,Donley2001,Jochim2003,Greiner2003,Zwierlein2005}. On the other hand, the detection methods for ultracold atoms have not changed very much since the beginning and absorption imaging is still the standard method to extract information from the system. The images are usually taken after a short time of flight in order to reduce the optical density and to magnify the atomic cloud. As a consequence the momentum distribution of the atoms is obtained rather than the spatial distribution inside the trap. \textit{In situ} measurements are possible using phase contrast imaging \cite{Andrews1996} and very small samples can be detected with absorption imaging even inside the trap \cite{Albiez2005}. However, all optical techniques inherently imply two limitations: they are not sensitive to single atoms and the spatial resolution is limited by the wavelength of the absorption light. Whereas the first can be overcome for very low atom numbers via flourescence imaging \cite{Schlosser2001,Kuhr2001} or optical cavities \cite{Pinkse2000,Oettl2005}, the latter constitutes a fundamental limitation. In fact, a spatial resolution of better than $2\,\mu$m is only achieved in few experiments \cite{Albiez2005,Schlosser2001}. This is especially relevant because the typical interatomic distances in ultracold quantum gases are on the order of a few hundred nanometers. It should be mentionend that in the case of metastable noble gas atoms, a powerful single atom detection technique based on micro channel plates has been developed \cite{Schellekens2005}. 

For solid objects the imaging resolution can be significantly improved by the use of electron microscopes. The small de Broglie wavelength of the electrons allows for a much better resolution whose ultimate limit is nowadays far below $1$\,nm. In this letter we discuss the use of scanning electron microscopy for the detection of single atoms in an ultracold quantum gas and show how both limitations can be overcome. Such a technique would be an important tool for the measurement of atomic correlations in quantum gases and for the study of the microscopic dynamics of atoms. In combination with optical lattices the method offers the possibility of single site addressing, which is a prerequisite for several proposals for quantum information processing in cluster states \cite{Raussendorf2000}.

The article is organized as follows: In section \ref{working} we introduce the working principle of the microscope.
Section \ref{interaction} gives an overview of the possible interaction mechanisms of electrons and neutral atoms with special emphasis on secondary reaction products and possible heating and loss mechanisms of the atomic ensemble. In the last section we briefly describe a possible experimental realization and comment on several technical and physical requirements that have to be fullfilled in order to make the technique work.

\section{Working principle}
\label{working}

In Fig.\,1 we show the working principle of the proposed technique. The focussed electron beam is directed onto the atomic sample. Atoms hit by an electron get ionized and are extracted with an electrostatic field towards the detector. Since the ions starting velocity is very small (see following section) the arrival time of the ion is unambigously connected to the position of the electron beam. Thus, the original position of the atom can easily be reconstructed from the detector signal. By scanning the beam across the cloud the spatial distribution of the atoms in the cloud can be imaged. Pointing the beam at a fixed position in the cloud allows one to measure \textit{in situ} the local density and temporal correlations between the atoms. In combination with optical lattices the electron beam can as well be used to remove atoms deterministically from specific lattice sites, thus providing a powerful preparation and detection tool for further experiments. The spatial resolution of the imaging technique is determined by the spot size of the electron beam and can be as small as $100$\,nm or even less, only depending on the required beam current. Besides the high resolution, the technique features a large depth of focus of a few tens of micrometers. Note, that the position of the atom along the electron beam axis cannot be reconstructed in this simple scheme and the imaging is two-dimensional. However, this can be overcome by implementing an ion optical extraction system which is followed by a spatially resolved detection of the ion. The position can then be reconstructed in the third dimension \cite{threedimensional} and the imaging technique becomes fully three-dimensional.

\begin{figure}[t]\label{hallo}
\vbox{\hbox to\hsize{\psfig{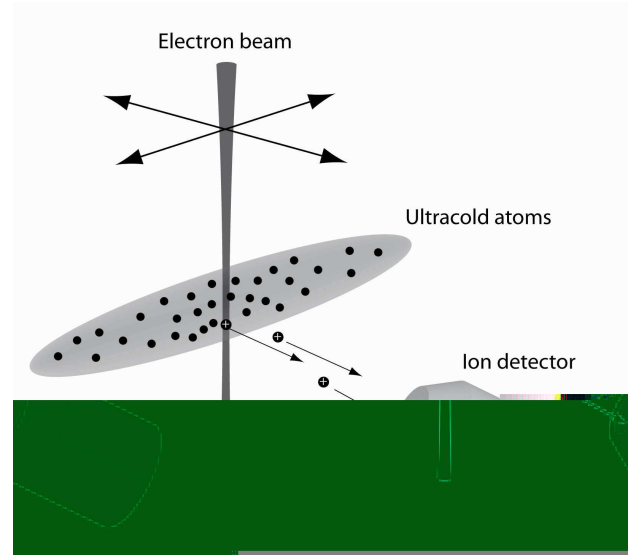}\hfill}}
\caption{Sketch of the working principle: The focussed electron beam is directed onto the atomic ensemble. Electron impact leads to the ionization of the atoms and the resulting ions are detected. The electron beam can be moved in two directions with deflectors.}
\end{figure}

\section{Electron-atom interaction}\label{interaction}

The interaction between fast electrons and atoms is goverened by three mechanisms: electron impact ionization, inelastic and elastic scattering. In this section we give quantitative estimates for the three interaction mechanisms and discuss their consequences on the proposed detection scheme. In the remainder we restrict ourselves to the collisional system electron-rubidium, though many results can be assigned to other alkali atoms.

We begin with the ionization processes which constitute the ''good'' events for the proposed technique. The total cross section for electron impact ionization of rubidium atoms has been experimentally determined by several groups \cite{Brink1964,McFarland1965,Chen1978,Wei1993,Keeler2000,Vuskovic1984}. The most precise value can be taken from Schappe et al. \cite{Schappe1995,Schappe1996} for an impact energy $E_i=500$\,eV and amounts to $2.5\times10^{-16}$\,cm$^2$. With a total scattering cross section of $6.2\times10^{-16}$\,cm$^2$ the ionization processes represent 40 percent of all scattering events. For higher impact energies this fraction is expected to remain constant \cite{Walker2004}. To quantify the typical time scale $t_i$ for the ionization of an atom located in the beam center we have to evaluate the following expression \cite{Schappe1995}
\begin{equation}\label{eq1}
t_i=\frac{q}{j_0}\times\frac{1}{\sigma_{ion}},
\end{equation}
where $j_0$ is the current density in the beam center, $q$ is the electron charge and $\sigma_{ion}$ is the total ionization cross section which depends on the electron energy. In expression (1) we have definded $t_i$ as the time in which the ''survival probability'' of the atom has dropped to $1/e$. At an electron energy of $E_i=5$\,keV modern electron columns can deliver $100$\,nA current into a $100$\,nm spot, resulting in a peak current density of $j_0=2.5\times 10^{3}$\,A/cm$^2$. Assuming a scaling of the ionization cross section proportional to $1/E_i$ \cite{Inokuti1971} we obtain $\sigma_{ion}=2.5\times10^{-17}$\,cm$^2$ and find $t_i=2.5\,\mu$s. A degenerate quantum gas has a typical density of $10^{14}$\,cm$^{-3}$ and a temperature of $T=100$\,nK. The gas is cigar-shaped with an axial width of several tens of micrometers and a radial extension of a few micrometers. The average velocity of the atoms is around $5$\,mm/s. From these parameters we can draw the following conclusions: (a) Most electrons pass the cloud and only 1 of $100,000$ electrons interacts with an atom. Thus, multiple scattering can be neglected. (b) The ionization probablity is high enough in order to ionize an atom before it moves out of the electron beam. (c) Only a small part of the cloud can be imaged simultaneously. For longer observation times the motion of the atoms has to be taken into account. (d) The depth of focus is larger than the diameter of the atomic cloud.

The dominant part of the ionization events takes place for small momentum transfers which are just sufficient to knock out the target electron \cite{Inokuti1971,Coplan1994}. The ion plays the role of a spectator and balances the total momentum. The target electrons initial velocity after ejection is large enough in order to escape from the atomic cloud without noticeable interaction with the remaining atoms. The much heavier ion is produced with a starting velocity of only a few m/s corresponding to the momentum distribution of the target electron. At these low kinetic energies, the elastic scattering cross section of the ion with the surrounding neutral atoms is very large and can exceed the s-wave scattering cross section for two neutral atoms \cite{Cote2000}. If the atomic ensemble is in the hydrodynamic regime and if the ion is produced inside the atomic cloud there is a possibility for a secondary reaction between the ion and the neutral atoms. However, the electrostatic field accelerates the ion very quickly, the elastic cross section drops and the cloud becomes more and more transparent. The magnitude of the electrostatic field is therefore a handle to influence the strength of secondary ion-atom collisions. Because traps for neutral atoms are very shallow most secondary reaction products have enough kinetic energy to escape from the trap.

The amount of inelastic scattering events can be calculated in first Born approximation and turns out to be on the same order of magnitude as the ionization channel. In the case of alkali atoms, the excitation is mainly on the strong s-p transition of the valence electron. The momentum transfer to the atom is small but large enough to kick the atom out of the trap. Inelastic scattering is not only a loss channel because the excited atom also emits a resonant photon. Since ultracold atomic ensembles are optically thick the emitted photon is reabsorbed with high probability. Each reabsorption process deposits in average one recoil energy in the cloud. As a consequence a heating rate is present. This effect can be supressed by an additional laser beam which ionizes the atoms from their excited state. It has the additional advantage that the overall ionization probability is enhanced and the detection efficiency is increased. The elastic scattering processes do not affect the internal state of the atom and solely momentum is transferred. They amount to 5-10 percent and constitute an additional loss channel that is difficult to avoid.
 
We finally calculate the static electric and magnetic fields of the electron beam for the parameters given above. The space charge of the electron beam creates an electrostatic field with a peak of $800$\,V/m at a distance of $40$\,nm from the beam center. For larger distances it falls off as $1/r$. The magnetic field of the beam amounts to $2$\,mG at a distance of $100$\,nm from the beam center and also falls off as $1/r$ for larger distances. Hence, the interaction of both fields with the ultracold atoms is extremely small and can be neglected outside the beam volume.    

\section{Experimental realization}\label{setup}

\begin{figure}[t]\label{fig2}
\vbox{\hbox to\hsize{\psfig{figure=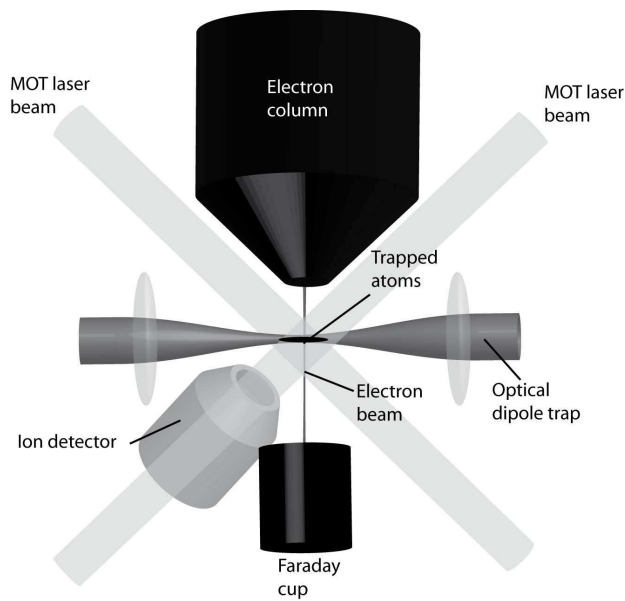,width=\hsize,clip=}\hfill}}
\caption{Experimental realization of the microscope: The electron column which produces the focussed electron beam is shown on top. With three pairs of counterpropagating laser beams (only two are shown) a magneto-optical trap is operated below the column tip. After collection in the magneto-optical trap, the atoms are transferred into an optical dipole trap. The ions are extracted with an electrostatic field and detected.}
\end{figure}

The combination of scanning electron microscopy and ultracold atoms imposes certain technical requirements. First, the electron column that delivers the focussed electron beam has to be compatible with an ultrahigh vacuum (UHV). Modern experiments on surface analysis with electron scattering have proven that there is no general restriction on the UHV compatibility of electron columns. Secondly, the trapping potential for the atoms should not influence the incoming electron beam. This suggests the use of optical dipole traps instead of magnetic traps. In Fig.\,2 we show a sketch of a possible experimental setup. The electron column is shown on top. Its conical shape allows for good optical accesss. The electron column can work either with an electrostatic or magnetic lens system. For magnetic lenses one has to take into account the soft iron pole pieces which distort external magnetic fields and the heat poduced from the coil for the lens excitation. Electrostatic lens systems are more favorable in this regard, but they have higher spherical aberrations. The highest achievable current density in the beam center is therefore smaller and the detection becomes slower. The position of the electron beam can be controlled with electrostatic or magnetic deflectors inside the column. The ultracold atoms are placed below the electron column and held in an optical dipole trap. The working distance between the tip of the column and the cloud can range from a few millimeters to several centimeters \cite{beamparameters}. A Faraday cup serves as a controlled beam stop. To load the atoms in the dipole traps one can either superimpose a magneto-optical trap or employ an optical tweezer. Because optical dipole traps have proven to be a fast and efficient route to Bose-Einstein condensation \cite{Cennini2003} the study of degenerate quantum gases is a natural application of such a setup. The ion detector together with a simple extraction electrode can be installed in any orientation with respect to the dipole trap axis. Modern ion detectors can reach detection efficiencies of more than 90 percent. Provided that the ionization fraction can be increased to 90 percent with the aid of optical post ionization of excited atoms, an overall detection efficiency of more than 80 percent might be achieved.

\section{Conclusion}\label{conclusion}

We have proposed a new imaging technique for ultracold atoms which is based on scanning electron microscopy. The technique is fast, highly efficient and can be extended to a fully three-dimensional detection scheme. With a dwell time of a few microseconds per pixel the technique develops its full capacity for small clouds or small parts of larger clouds. The spatial resolution is determined by the diameter of the electron beam which can be made so small that it represents no limitation. The technique can be used to study spatial and temporal correlation functions of ultracold atoms. In combination with optical lattices, it also offers the unique possibility for single site addressing of individual atoms.

\begin{acknowledgement}
We would like to thank E. Plies, C. Zimmermann, T. Kirchner, C.D. Lin, I. Bloch and H. Schmidt-B{\"o}cking for valuable discussions. We gratefully acknowledge financial support from the DFG under Grant No. Ot 222/2-3.
\end{acknowledgement}


\begin{thebibliography}{10}
\bibitem{threedimensional} The spatial resolution along the electron beam axis depends crucially on the starting velocity of the ions and the extraction field geometry. Simulations show that a resolution of $1\,\mu$m might be possible.
\end{thebibliography}
\end{document}